\documentclass[aps,prd,reprint,superscriptaddress]{revtex4-1}
\usepackage{graphicx}
\usepackage{amssymb}
\usepackage{amsmath}
\usepackage{bm}
\usepackage{verbatim}

\usepackage{enumitem} 

\newcommand{\der}{\partial}

\begin{document}

\title{Critical velocity of a mobile impurity
 in one-dimensional quantum liquids}

\author{M. Schecter}
\affiliation{School of Physics and Astronomy, University of Minnesota,
Minneapolis, MN 55455, USA}

\author{A. Kamenev }
\affiliation{School of Physics and Astronomy, University
of Minnesota, Minneapolis, MN 55455, USA}
\affiliation{William I. Fine
Theoretical Physics Institute, University of Minnesota, Minneapolis, MN
55455, USA}

\author{D.M.~Gangardt}
\affiliation{School of Physics and Astronomy, University of Birmingham,
Edgbaston, Birmingham, B15 2TT, UK}

\author{A.~Lamacraft}
\affiliation{Department of Physics, University of Virginia,
Charlottesville, Virginia, 22904-4714, USA}

\date{\today}

\begin{abstract}
We study the notion of superfluid critical velocity in one spatial
dimension. It is shown that for heavy impurities with mass $M$ exceeding
a critical mass $M_\mathrm{c}$, the dispersion develops periodic
metastable branches resulting in  dramatic changes of dynamics in the
presence  of an external driving force. In contrast to smooth Bloch
Oscillations for $M<M_\mathrm{c}$, a heavy impurity climbs metastable
branches until it reaches a branch termination point or undergoes a
random tunneling event, both leading to an abrupt change in velocity and
an energy loss. This is  predicted to lead to a non-analytic dependence
of the impurity drift velocity on small forces.
\end{abstract}
\maketitle

The study of the effect of impurities constitutes a major theme in condensed
matter physics. In the solid state impurities are typically regarded as
\emph{immobile}, acting as static perturbations.  Prime examples include the
residual resistance of metals and the Kondo effect. In other (typically
liquid) states \emph{mobile} impurities play a vital role, with the influence
of $^{3}$He impurities on the transport properties of superfluid $^{4}$He
being perhaps the best studied example
\cite{LandauKhalatnikov1949ViscosityI,*LandauKhalatnikov1949ViscosityII,
  BaymEbner1967Phonon} .

The traditional goal has been to understand the effect of foreign
particles or defects on an otherwise pure system. However, the
converse problem -- to determine the effect of the surrounding
medium on the motion of impurities -- is increasingly relevant to
the study of ultracold atomic gases. There are several ways in
which mobile impurities appear naturally in these systems: (i) a
small fraction of atoms may be transferred to a different
hyperfine state, creating a dilute gas of distinguishable
impurities of the same mass
\cite{Ketterle_impurity_scattering2000,
Zwierlein_fermi_polarons2009,PhysRevLett.103.150601}; (ii)
mixtures of different atomic species (\emph{e.g.}, $^{87}$Rb and $^{41}K$
in Ref.~\cite{Catani_K_Rb_mixtures2008}) with one dilute component
give a route to impurities of mass $M$ different from the mass $m$
of the host particles; (iii) \emph{Ions} of $\mathrm{Yb}^+$,
$\mathrm{Ba}^+$ or $\mathrm{Rb}^+$ have been placed in a
Bose-Einstein condensate of neutral ${}^{87}\mathrm{Rb}$ atoms
\cite{Kohl_trapped_ion2010,Denschlag_trapped_ion2010}.  In each of
these situations, impurity dynamics may be investigated with the
help of external forces that act selectively on the impurity.

\begin{figure}[!h]
\includegraphics[width=\columnwidth]{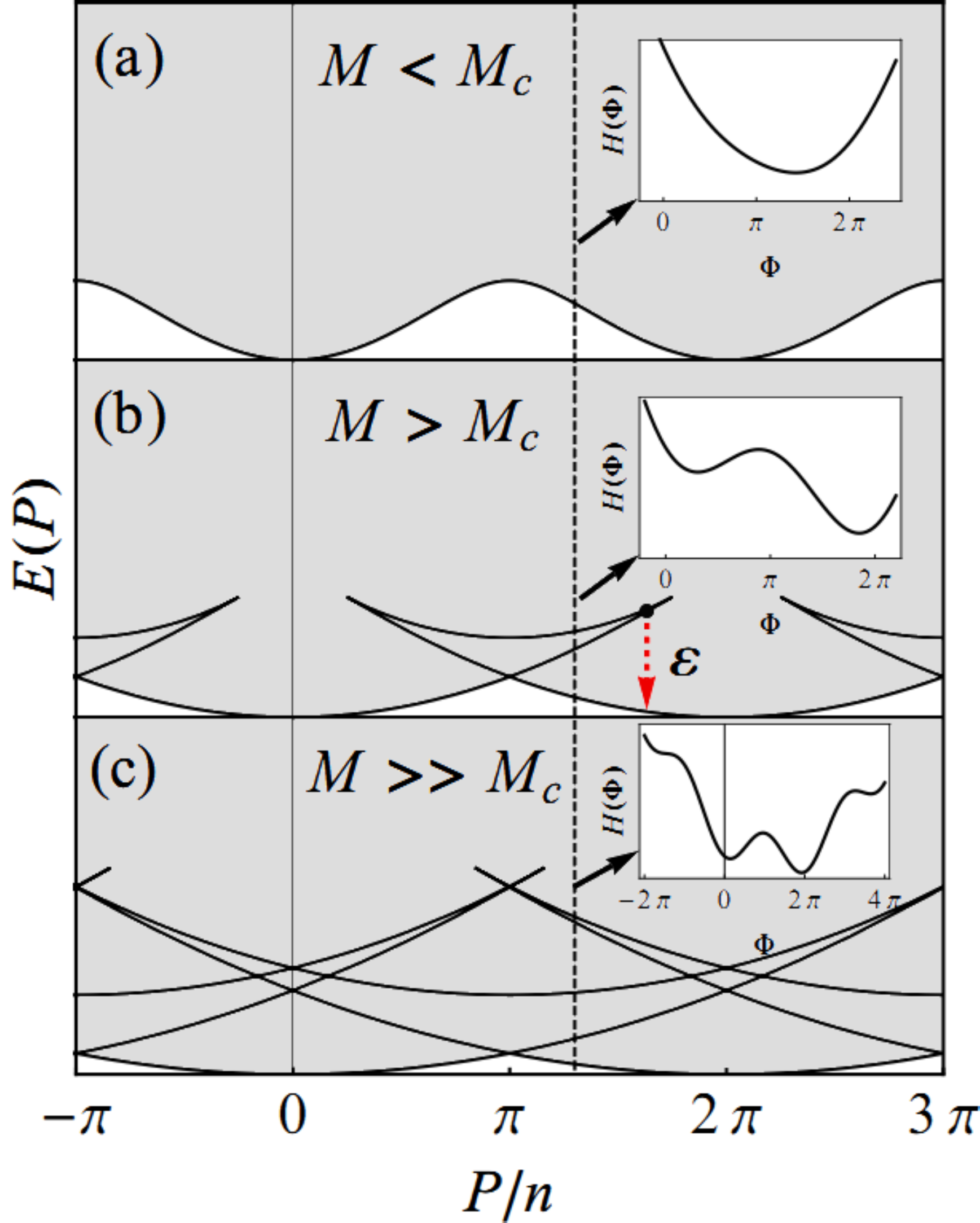}
\caption{Dispersion law for an impurity in a 1D quantum fluid for various impurity masses. \textbf{(a)} $M<M_c$, $E(P)$ is a smooth function of $P$ and $H$ (inset, Eq.~(\ref{eq:JJplusB})) has a single minimum. \textbf{(b)} $M\gtrsim M_c$, there exists a metastable minimum. An impurity driven by a force climbs a metastable branch until tunneling or $H$ loses its minimum. Each cycle releases energy $\varepsilon$ to the system. \textbf{(c)} $M\gg M_c$, many minima co-exist.}
\label{fig:dispersion}
\end{figure}

Owing to recent advances in cold atom trapping, systems in reduced dimensionalities are now well within experimental reach \cite{PhysRevLett.103.150601,Kohl_trapped_ion2010,Denschlag_trapped_ion2010}, bringing to light a wide range of rich and peculiar phenomena. In one spatial dimension (1D) Refs.~\cite{zvonarev_2007,akhanjee_2007} established that even in the limit of strong interactions, there exists a timescale beyond which an impurity with quadratic dispersion, $E(P)=P^2/2M^*$, may propagate in a quantum fluid. In this Letter we focus on the full dispersion $E(P)$ of the resulting motion, which possesses several remarkable features not present in higher dimensions. 
The most prominent property is its \emph{periodicity}, which is not evident from a quadratic, small $P$ expansion, and may be explained as follows. The motion of $N$ particles on a
ring with circumference $L$ results
in the \emph{total} momentum being quantized in units of
$2\pi \hbar N/L = 2\pi\hbar n$, where $n$ is the
one-dimensional density of the fluid. On the the other hand, in the thermodynamic limit $N,L\to \infty$, $N/L=n$, this motion costs no kinetic energy due to infinite total mass $Nm$.  Thus, instead of paying macroscopic energy $(2\pi\hbar n)^2/2M^*$ to deliver this momentum to impurity motion, the system channels it to the superflow of the background liquid at no energy cost. As a result, the dispersion of the dressed impurity is a periodic function $E(P)=E(P+2\pi\hbar n)$. Owing to the quadratic dispersion near $P=0$, impurity motion costs less energy for a given momentum than the softest excitations of the liquid, \emph{i.e.,} phonons with linear dispersion. This fact implies that impurity motion, while possessing a periodic dispersion, also defines the lower edge of the many-body continuum, above which (shaded grey, Fig.~\ref{fig:dispersion}) there exist excitations comprised of the moving impurity and phonons \cite{zvonarev_2009,imam_glaz_PRL2009,kam_glaz_PRA2009,Lamacraft_impurity_dispersion2009}. Explicit examples  demonstrating these features are furnished by models solvable via Bethe Ansatz \cite{Yang1966OneDimensional,Gaudin_1967,Castella_Zotos_1993}. A dramatic consequence of the dispersion periodicity is the possibility of observing Bloch oscillations of a driven impurity in 1D \emph{in the absence of a lattice} \cite{Gangardt09,Schecter_bloch_oscillations2011}.

The same picture of momentum transfered to
the fluid at no energy cost leads to the
prediction of vanishing  Landau critical velocity in 1D
\cite{Buechler_Geshkenbein_Blatter_2001,Astrakharchik2004Motion}.
It was shown that an \emph{infinitely massive}
object moving with any nonzero
velocity with respect to the background nucleates phase slips
corresponding to one quantum $2\pi \hbar n$ of total momentum
transferred to the fluid. The finite rate of this process can then be
related to the energy and momentum dissipation in the liquid, precluding
superfluidity. On the other hand, a mobile impurity with \emph{finite mass}
$M$ can propagate without dissipation at zero temperature with velocity
given by slope of the dispersion $V=\der E/\der P$. Hence,
the maximum slope of the dispersion defines a non-zero critical velocity for
light mobile impurities.

The main result of this work is the prediction that the above
regimes of heavy and light impurities are separated by a quantum
phase transition taking place at a critical mass $M_\mathrm{c}$,
given by Eq.~(\ref{eq:mcrit}). We show that above the critical
mass the many-body ground state in a total momentum sector $P$ has
cusps at $P=\pm \hbar \pi n, \pm 3\hbar \pi n,\ldots$, while the
impurity dispersion develops a swallowtail structure with
metastable branches shown in Fig.~\ref{fig:dispersion}. In this
case the true {\em thermodynamic} critical velocity
$\mathcal{V}_\mathrm{c}$, given by the maximal slope of the
many-body ground state, should be distinguished from the {\em
dynamic} critical velocity $V_\mathrm{c}$ obtained from the
maximal slope of the metastable branch.
It can be shown that for $M\gg M_c$ the thermodynamic critical
velocity is inversely proportional
to  $M$ consistent with  the infinite mass limit of
Refs. \cite{Buechler_Geshkenbein_Blatter_2001,Astrakharchik2004Motion}.

\emph{Metastability and two definitions of critical velocity}.  For a concrete
example of how a metastable impurity dispersion can arise, let us take our
fluid to be a weakly interacting Bose gas.  At the mean field level,
appropriate for weak interactions, the condensate wavefunction develops a
phase drop $\Phi$ across the impurity located at $X$.  In the limit of strong
coupling between gas and impurity, the latter gives rise to the Josephson term
\begin{eqnarray}
  \label{eq:JJ}
  H_\mathrm{d} (\Phi ) =-n {V}_\mathrm{c}\cos\Phi
\end{eqnarray}
(we have set $\hbar =1$) in the impurity energy. Here
${V}_\mathrm{c}$ is the dynamical critical velocity which depends
on the details of the impurity. The phase drop inevitably creates
a background supercurrent contributing a term $n\Phi$ to the {\em
total} momentum $P$. The total energy of the impurity and the
background can then be written as
\begin{eqnarray}
  \label{eq:JJplusB}
  H (P,\Phi) = \frac{1}{2\mathcal{M}} (P-n\Phi)^2 +H_\mathrm{d}(\Phi)
\end{eqnarray}
The first term is the kinetic energy of the impurity moving with velocity
$V=(P-n\Phi)/\mathcal{M}$, where $\mathcal{M}$ is the total mass of the
impurity and induced depletion cloud. As we shall see one can assume
$\mathcal{M}\simeq M$ for a sufficiently heavy impurity.

The phase drop $\Phi$ represents a collective coordinate
characterizing the state of the impurity equilibrated with the
background \cite{Schecter_bloch_oscillations2011}.  Its value is
determined from the requirement of the minimum of the total energy
(\ref{eq:JJplusB}) leading to the matching $P-n\Phi =
M{V}_\mathrm{c}\sin\Phi$ of the Josephson current $n
{V}_\mathrm{c}\sin\Phi$ to the current $nV$ experienced by the
impurity in its rest frame.  For sufficiently large mass
$M>M_\mathrm{c}$ this equation has several solutions corresponding
to different arrangements of the impurity and the background.  The
total energy $H(P,\Phi)$ in Eq.~(\ref{eq:JJplusB}) can therefore
have minima for several values of the phase drop $\Phi$ for a
given value of the total momentum $P$.  Plotting the corresponding
energies results in a typical swallowtail structure for the
dispersion, see Fig.~\ref{fig:dispersion}.

Following the
global minimum $E_- (P)$ of the dispersion  one sees that it develops cusps at the
crossing points $P=\pm \pi n,\pm 3\pi n,\ldots$ corresponding to a first order
transition as a function of the total momentum $P$.
In addition to the true ground state $E_- (P)$ there is also a metastable
branch $E_+ (P)$ corresponding to local minimum. The minima corresponding to
$E_- (P) $ and $E_+ (P)$ are separated by the maximum.  At the termination
points $P_\mathrm{t}$ the maximum merges with the local minimum at $\Phi =
\Phi_\mathrm{t}$ and the metastable branch ceases to exist.
The critical mass $M_\mathrm{c}$ for the swallowtail catastrophe
can be estimated from the simple model Eq.~(\ref{eq:JJplusB}) as
\begin{eqnarray}
  \label{eq:mcrit}
 M_\mathrm{c} = \pi n/{V}_\mathrm{c} =K m ({c}/{V}_\mathrm{c})\ ,
\end{eqnarray}
where $K=\pi n/mc$ is the Luttinger parameter of the liquid
depending on the speed of sound $c$. For a strongly repulsive
impurity in a weakly interacting background $M_c \gg m$ since
$K\gg 1$ and $c \gg {V}_\mathrm{c}$, justifying \emph {a
posteriori} the assumption  $\mathcal{M}\simeq M$.

For $M\gg M_\mathrm{c}$ the termination point corresponds to
$\sin\Phi_\mathrm{t}=\pm 1$ and the mass-independent velocity
$\der E_+/\der P = \pm {V}_\mathrm{c}$. In contrast, the
thermodynamic  critical velocity, defined with the help of the
stable branch $\mathcal{V}_\mathrm{c} =\der E_-/\der P$, behaves
as $1/M$ resulting in zero critical velocity for infinitely heavy
impurity
\cite{Buechler_Geshkenbein_Blatter_2001,Astrakharchik2004Motion}.

Quantum fluctuations can smear the transition  by producing a
mechanism for the decay of the  metastable branch
\cite{Lamacraft_impurity_dispersion2009}. The metastable branch
$E_{+}(P)$ is well defined as long as the decay rate $\Gamma(P)$
to the lower branch satisfies $\Gamma(P)\ll
E_{+}(P)-E_{-}(P)\approx 2\mathcal{V}_\mathrm{c}|P-\pi n|$. We
show below that $\Gamma(P)$ vanishes as a power law near the
crossing points, $\Gamma(P) \propto |P-\pi n|^{\alpha}$ for
$P\to\pi n$ with the exponent $\alpha$ given in
Eq.~\eqref{eq:rate1}. Self-consistency therefore requires
$\alpha>1$. Hence $\alpha=1$ is the condition defining the transition
between the two classes of dispersion curves, and corresponds to $M=M_c$
\cite{Lamacraft_impurity_dispersion2009}.

Before proceeding we note that the qualitative features discussed
above in no way depend upon the specific model Eq.~\eqref{eq:JJ}
and occur (for example) in the dispersion relation of a particle
moving in a Bose gas described by the Gross--Pitaevskii equation,
valid for arbitrary coupling between the gas and impurity \cite{sup_mat}.

\emph{Depleton model}. We now introduce a framework to describe the periodic
impurity dispersion, the metastable branch, and its decay in the general case.
The moving depleton, \emph{i.e.} impurity dressed by the liquid depletion, is
characterized by a number of particles $N$ expelled from its core in addition
to the phase drop $\Phi$. The existence of the two collective slow coordinates
is associated with the presence of the two conservation laws: number of
particles and momentum~\cite{Schecter_bloch_oscillations2011}.  The depleton's
dynamics is then specified by the following Hamiltonian
\begin{equation}
\label{eq:ham}
H(P,\Phi,N)=\frac{1}{2\mathcal{M}}(P-n\Phi)^2+\mu N+H_{\mathrm{d}}(\Phi,N).
\end{equation}
Here $H_{\mathrm{d}}$ is the $\Phi$-periodic energy function of
the depletion cloud, $\mu$ is the chemical potential of the liquid
in the absence of the impurity and $\mathcal{M} = M-mN$. In the
limit of strongly repulsive impurity the dynamics of $N$ is frozen
and $H_{\mathrm{d}} (\Phi,N)$ reduces to the simple Josephson form
Eq.~(\ref{eq:JJ}).  If the mass $\mathcal{M}$ is sufficiently
large, the Hamiltonian (\ref{eq:ham}) (being almost a periodic
energy function of $\Phi$)  possesses many metastable minima in
addition to the absolute one.

We wish to determine the tunneling rate from the above mentioned
metastable minima to the ground state branch at the same momentum
$P$. Such tunneling is accompanied by a change in the phase drop
$\Delta\Phi$ and number of depleted particles $\Delta N$, despite
the momentum of both states being equal \cite{Note2}. Therefore one must know the
Lagrangian governing the dynamics of the collective variables $\Phi$ and
$N$. As shown in Ref.~\cite{Schecter_bloch_oscillations2011}, this requires
the introduction of, and coupling to, the phonon subsystem. The latter may be
described by small deviations of the density, $\rho(x,t)$, and velocity,
$u(x,t)$, fields from their unperturbed values. It is convenient to introduce
the superfluid phase $\varphi(x,t)$ and the displacement field
$\vartheta(x,t)$ defined by $u = \der_x \varphi/m$ and $\rho = \der_x
\vartheta/\pi$.

The effective action governing these harmonic degrees of freedom is that
of the Luttinger liquid \cite{PopovBookFunctional,HaldanePRL81}
\begin{equation}
    \label{eq:LL_Ham}
S_{\text{LL}} = \frac{1}{\pi}\int \mathrm{dt}\mathrm{dx}
\left[-\der_x\vartheta\der_t\varphi
-\frac{c}{2K}(\partial_x\vartheta)^2-\frac{cK}{2}(\partial_x\varphi)^2\right],
\end{equation}
while their coupling to the depleton degrees of freedom takes a
universal form \cite{Schecter_bloch_oscillations2011,Note4},
\begin{equation}
\label{eq:phonon_action} S_\mathrm{int}=\int \mathrm{d}t \left[ -
\dot \Phi\; \vartheta(X,t)/\pi -\dot N\varphi(X,t)\right].
\end{equation}
It is the phonon subsystem which must supply the macroscopic
momentum $n\Delta\Phi$ and take away $\Delta N$ particles from the
depleton. Near the point $P=\pi n$, the phonons have only a small
amount of energy $\varepsilon$ with which to do this. This low-probability event, or `under-the-barrier' evolution of the
phononic system, can be described in imaginary time by
performing the Wick rotation $t\rightarrow-i\tau$ in
Eqs.~(\ref{eq:LL_Ham}), (\ref{eq:phonon_action}). We have delegated details of the tunneling calculation to a short section in the Supplemental Material \cite{sup_mat}. We find for the tunneling rate
\begin{equation}
\label{eq:rate1}
\Gamma(\varepsilon)\sim\varepsilon^\alpha;\,\,\alpha=2K\left[\left(\frac{\Delta\Phi}{2\pi}\right)^2+\left(\frac{\Delta N}{2K}\right)^2\right]-1.
\end{equation}
Writing $\varepsilon=2\mathcal{V}_\mathrm{c}\left|P-\pi n\right|$ gives the
momentum dependent tunneling rate advertised earlier.

The logarithmic behavior of the tunneling action \cite{sup_mat} is valid only on
the long time-scale $\tau\gtrsim l/c$, implying the validity
of Eq.~(\ref{eq:rate1}) is restricted to the regime of small
$\varepsilon$ close to $P=\pi n$. In such a case the velocities of
the upper and lower branches of the dispersion are essentially
opposite and close to $\pm \mathcal{V}_\mathrm{c}$, while the
change in number of particles $\Delta N$, being an even function
of velocity, is negligible. As a result one arrives at the
exponent $\alpha=2K(\Delta\Phi/2\pi)^2-1$. This single collective
coordinate limit, \emph{i.e.} $\Delta N=0$, agrees with the
previous result of Ref.~\cite{Leggett_quantum_dissipation1987}
(cf. Eq.~(7.20) therein) based on a phenomenological model of
tunneling between two minima in the presence of coupling to an
environment. Their results also show that a finite temperature $\mathcal{T}$ acts as a low energy cut-off, namely $\Gamma\sim \mathcal{T}^\alpha$, $\varepsilon\ll\mathcal{T}$ (see Eqs.~(5.30), (7.18) of Ref.~{\cite{Leggett_quantum_dissipation1987}}).

Higher order processes involve tunneling back and forth between the two minima
that become degenerate for $\varepsilon=0$, leading to an effective spin-boson
or anisotropic Kondo description \cite{Lamacraft_impurity_dispersion2009}. The
inclusion of such processes is only necessary if $\alpha<1$, meaning that
tunneling is a relevant perturbation. This leads to the rounding out of the
cusp in the dispersion $E_{-}(P)$ at $P=\pi n$, and the disappearance of the
metastable branch. Thus $\alpha=1$ is the exact condition determining the
critical mass $M=M_{c}$.

In the infinite mass limit
with fixed velocity $V=P/M$  Eq.~(\ref{eq:ham}) becomes
essentially a washboard potential, Fig.~\ref{fig:dispersion}c, $
H\underset{M\rightarrow\infty}{\longrightarrow}nV\Phi+H_{\mathrm{d}}(\Phi)$
(change $\Delta N$ between adjacent minima goes to zero in this
limit). Thus the change $\Delta\Phi$ between the adjacent minima
is exactly $2\pi$, while the energy difference is
$\varepsilon=2\pi n V$. As a result, one finds $\alpha = 2K-1$ and
the power law dependence of the tunneling rate $\Gamma\sim
V^{2K-1}$ \cite{Note3} on the velocity in full agreement with the result of
Ref.~\cite{Buechler_Geshkenbein_Blatter_2001,Astrakharchik2004Motion}.

In the limit of large energy detuning, \emph{i.e.}, away from
$P=\pi n$, the tunneling rate does not take the form
(\ref{eq:rate1}). In this region the barrier is small and the
impurity emerges from under it with only slightly less energy,
despite the large detuning between the branches (see
Fig.~\ref{fig:dispersion} insets). Therefore most of the descent
from the upper branch corresponds to viscous relaxation in real
time and not an under-the-barrier tunneling trajectory. Thus, near
the termination momentum $P_\mathrm{t} \simeq MV_\mathrm{c}$, the
potential may be approximated by a cubic polynomial.  The
dissipative tunneling through such a potential was described in
Ref.~\cite{Caldeira1983Quantum}, leading to the escape rate
$\Gamma\propto e^{-\bar{S}}$, where
\begin{equation}
\label{eq:action}
\bar{S}=
2K\frac{P_\mathrm{t}-P}{\sqrt{M^2 V_\mathrm{c}^2-n^2}}.
\end{equation}
We see that for both large and small energy detuning, Eqs.~(\ref{eq:rate1}), (\ref{eq:action}) the tunneling action scales
with the Luttinger parameter $K$.

\begin{figure}[t] \centering
  \includegraphics[width=\columnwidth]{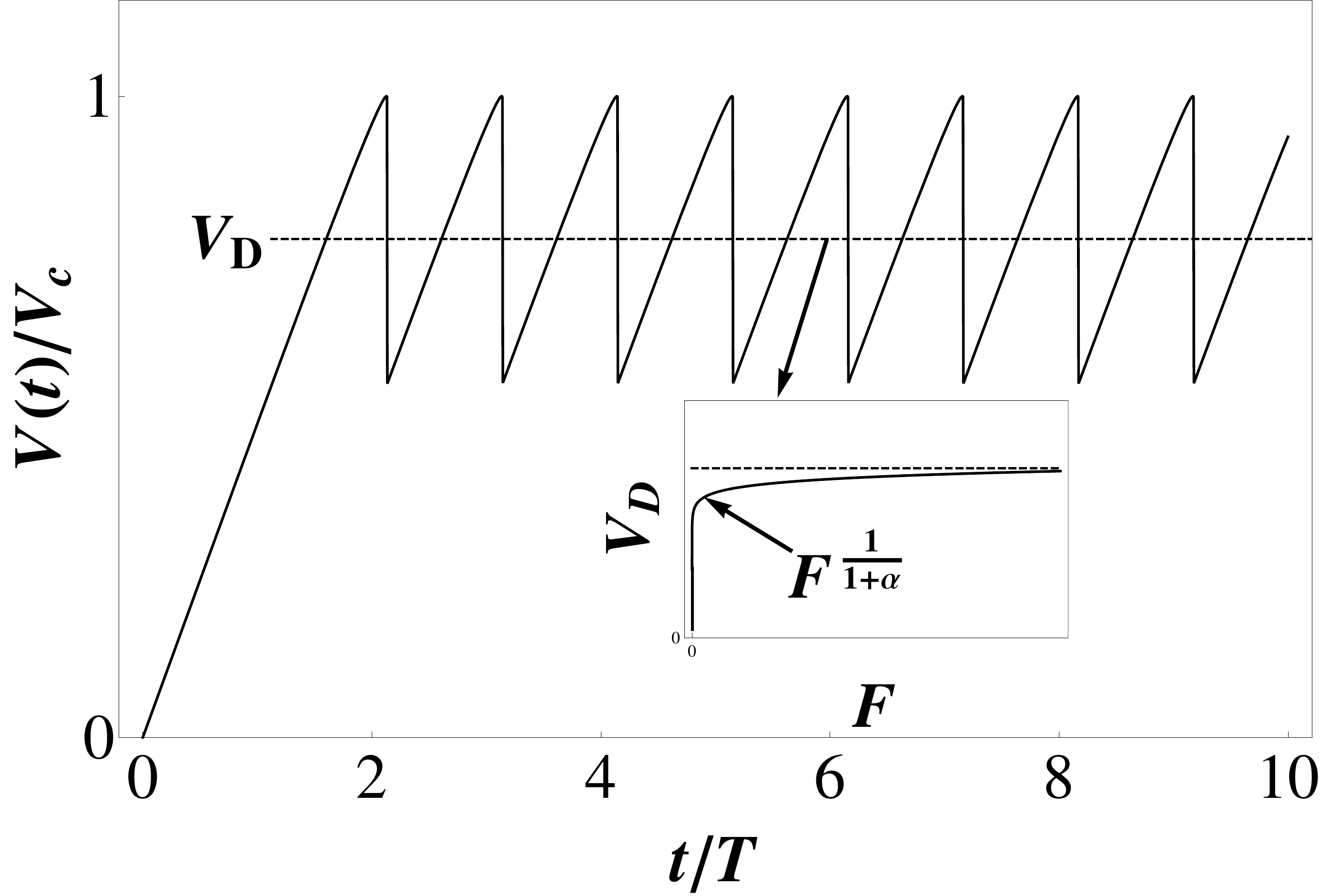}
  \caption{Impurity velocity (solid) in units of the dynamic critical velocity $V_{\mathrm{c}}$ as a function of time in units of the period $T=2\pi n/F$. The drift velocity
    is denoted by $V_{\mathrm{D}}$ (dashed) and is displayed in the inset as a function of the applied
    force. Due to quantum tunneling near $P=\pi n$, the drift velocity goes to
    zero as a power law with exponent $1/(\alpha+1)$ (solid, inset) given by
    Eq.~(\ref{eq:drift1}) for $F<F_\mathrm{c}$ and approaches the semi-classical
    result for $F>F_\mathrm{c}$ (dashed, inset) \cite{sup_mat}.}
\label{fig:velocity}
\end{figure}

One way to explore the metastable branches of the dispersion is by applying an
external force $F$ to the impurity and studying the ensuing dynamics.
For sufficiently strong force the driven impurity overshoots the ground state
branch and follows the metastable branch until it either tunnels or reaches
the termination point $P_\mathrm{t}$ (see Fig.~\ref{fig:dispersion}). The
energy dissipated per cycle, $\varepsilon$, must be supplied by external
force: $FV_{\mathrm{D}}=\varepsilon/T$ resulting in the drift velocity
$V_{\mathrm{D}}=\varepsilon/2\pi n$. Here $T=2\pi n/F$ is the period of the
motion.

Tunneling is negligible when $\Gamma T\ll 1$, which occurs if the force is
sufficiently large. In such a case the external force drives the impurity all
the way up to the termination point $P_\mathrm{t}$. It then loses the energy minimum
and relaxes to the lower branch, emitting phonons. As a result, the energy
dissipated per cycle and thus drift velocity are essentially independent of
the applied force.
This is in stark contrast to the case of $M<M_c$, where the phonon
emission due to dipole radiation gives rise to a linear in $F$
drift velocity and hence a \emph{finite} impurity mobility $\sigma
= V_\mathrm{D} /F$ \cite{Schecter_bloch_oscillations2011}.

The impurity velocity takes the form of a saw-tooth trajectory, experiencing a
slow build-up and a sudden drop when the velocity reaches $V_\mathrm{c}$, see
Fig.~\ref{fig:velocity}. As a function of the mass, $V_{\mathrm{D}}$ saturates
to the critical velocity in the limit $M\rightarrow\infty$. This is most
easily seen by noting that in such a limit the energy drop occurs between two
adjacent and essentially parallel parabolas: $\varepsilon=2\pi n V_\mathrm{D}
= P^2/2M-(P-2\pi n)^2/2M \to 2\pi nV_\mathrm{c}$. On the
same grounds, the amplitude of the saw-tooth oscillations is a decreasing
function of $M$ going to zero in the infinite mass limit.

Quantum effects become dominant only if the time spent on a
metastable branch is comparable with the inverse tunneling rate:
$\Gamma T\gtrsim 1$. Thus, the effect of quantum tunneling plays a
role only for very small forces. In this limit, the impurity is
likely to tunnel at a momentum immediately above $P=\pi n$, right
when it enters the metastable branch. Using the rate
Eq.~(\ref{eq:rate1}) to calculate the tunneling-averaged energy
drop $\langle\varepsilon\rangle$, we find for the drift velocity \cite{sup_mat}
\begin{equation}
\label{eq:drift1} V_{\mathrm{D}} = \frac{c}{K}\left(\frac{F}{F_\mathrm{c}}\right)^{\frac{1}{\alpha+1}}.
\end{equation}
The non-analytic dependence of the
drift on small forces is the result of the cusp of the ground
state energy at $P=\pi n$ and implies the divergence of the
mobility $\sigma=V_{\mathrm{D}}/F$ as $F\rightarrow 0$ across the
$M=M_c$ transition (see inset of Fig.~\ref{fig:velocity}). At finite temperature $\mathcal{T}$ the mobility is \emph{not} given by $\sigma\sim \mathcal{T}^{-\alpha}$ as might be expected due to the finite tunneling rate $\Gamma\sim\mathcal{T}^\alpha$ near the cusp. Strictly speaking, it is given by $\sigma\sim \mathcal{T}^{-4}$ due to two-phonon thermal Raman scattering \cite{Gangardt09,Schecter_bloch_oscillations2011,castro_fisher} which results in velocity saturation \emph{below} $\mathcal{V}_\mathrm{c}$ as $F\to0$. In practice, therefore, it is necessary to work at sufficiently low temperatures and finite drives to overcome such saturation and allow the impurity to explore the full range of momentum where metastable branches become important.

In conclusion, we have shown that the dispersion relation of a
mobile impurity undergoes a sudden change when the mass of the
impurity exceeds a certain critical value. The latter depends on
the interaction parameter within the host liquid and thus may be
varied experimentally. The predicted transition can be probed by
applying an external force and monitoring either impurity velocity
or excitations of the host liquid. The change of dispersion also
affects the oscillation frequency and dissipation of the impurity
in the trapping potential. With investigations of impurity motion
in 1D atomic gases now underway
\cite{PhysRevLett.103.150601,Kohl_trapped_ion2010,Denschlag_trapped_ion2010}, we hope
that these predictions will guide future experiments.

This work was supported by the EPSRC Grant No.\ EP/E049095/1. MS
and AK were supported by DOE contract DE-FG02-08ER46482.   DMG
acknowledges support by EPSRC Advanced Fellowship EP/D072514/1. AL
acknowledges support of the NSF grant DMR-0846788 and the Research
Corporation through a Cottrell Scholar award.

\section{Supplemental Material}

\emph{Impurity coupled to a weakly interacting liquid} -- To model the heavy impurity in the presence of a weakly interacting superfluid at $T=0$, we employ
the Gross-Pitaevskii equation (GPE) using a point description of the impurity wavefunction with coordinate $X$,
\begin{equation}
i\partial_{t}\Psi=\left(-\frac{\partial_{x}^{2}}{2m}+
g|\Psi|^{2}-\mu+G\delta(x-X)\right)\Psi.
\label{eq:gross pitaevskii}
\end{equation}

We look for a solution of the traveling wave form $\Psi(x,t)=\Psi(x-Vt)$ and $X(t)=Vt$. The usual way of dealing with a delta function perturbation is to attempt a solution constructed from two homogeneous (\emph{i.e.}, $G=0$) solutions that must satisfy the proper boundary conditions at $x=X$. This strategy is facilitated by the fact that the bare GPE ($G=0$) admits a one-parameter family of solutions
\begin{equation}
\Psi_\mathrm{s}(x-Vt)=
\sqrt{n}\left(\frac{V}{c}-i\sqrt{1-\frac{V^2}{c^2}}\textrm{tanh}
\left(\frac{x-Vt}{l}\right)\right),\label{eq:wavefunction_no_impurity}
\end{equation}
known as grey solitons \cite{PitaevskiiStringariBook,Tsuzuki_1971}.  They can
be visualized as a density dip moving with velocity $V$, having a core size
$l=\xi(1-V^2/c^2)^{-1/2}$, where $c=\sqrt{ng/m}$ is the speed of sound, $n=\mu/g$ is the 1D density and $\xi=1/mc$ is the healing length (the condition of weak inter-particle coupling implies $n\xi\gg1$). Thus, by appropriately matching two solitonic solutions at $x=Vt$, one formally solves Eq.~(\ref{eq:gross pitaevskii}) with $\Psi(x-Vt)$ given by
\begin{eqnarray}
\Psi(y)=\begin{cases}
\Psi_\mathrm{s}(y+x_{0})e^{i\Phi_{0}/2}, & y>0\\
\Psi_\mathrm{s}(y-x_{0})e^{-i\Phi_{0}/2}, & y<0\end{cases}
\label{eq:psi_with_impurity}
\end{eqnarray}
where $y=x-Vt$. Here we introduced two "matching parameters" $x_0$ and $\Phi_0$ which must be chosen to satisfy the two boundary conditions: $\Psi(0^+)=\Psi(0^-),\,\Psi^\prime|^{x=0^+}_{x=0^-}=2mG\Psi(0)$. This results in the
following two equations for $\Phi_{0}$ and $z=\textrm{tanh}(x_{0}/l)$
\begin{eqnarray}
\label{eq:z1}
\tan\frac{\Phi_{0}}{2}&=&z\,\tan
\frac{\Phi_{s}}{2},\\
\label{eq:z2}
\sin^{3}\frac{\Phi_\mathrm{s}}{2}\left(1-z^{2}\right)z
&=&\frac{G}{c}\left[\cos^{2}\frac{\Phi_\mathrm{s}}{2}
+z^{2}\sin^{2}\frac{\Phi_\mathrm{s}}{2}\right],
\end{eqnarray}
where we have parameterized the velocity $V$ by the solitonic phase $\Phi_{\mathrm{s}}$ via $V/c=\mathrm{cos}\frac{\Phi_{\mathrm{s}}}{2}$.

Equations (\ref{eq:z1}),~(\ref{eq:z2}) permit a solution only for
$V<V_{c}$ where $V_{c}=V_c(G)<c$ is the dynamic critical velocity that depends only upon
the parameter $G$
\cite{Hakim_PhysRevE.55.2835,Gunn_Taras_PhysRevB.60.13139}. This may be seen
by plotting the right and left hand sides of Eq.~(\ref{eq:z2}). While the left hand side is bounded by a
maximum, the right hand side grows quadratically
with $z$ and therefore the solution exists only for a limited range of $\Phi_s$. For this reason we choose
to parameterize the solution, Eq.(\ref{eq:psi_with_impurity}), by the
\emph{total} phase drop across the impurity, $\Phi=\Phi_{s}-\Phi_{0}$, which permits a solution for \emph{any} $\Phi$. Upon solving Eqs.~(\ref{eq:z1}),~(\ref{eq:z2}) one thus finds the periodic relations
$z=z(\Phi,G/c)$ and $\Phi_\mathrm{s}=\Phi_\mathrm{s}(\Phi,G/c)$ (\emph{i.e.}, they are invariant with respect to $\Phi\to\Phi+2\pi j$ for integer $j$).

\begin{figure}[t]
\centering
\includegraphics[width=\columnwidth]{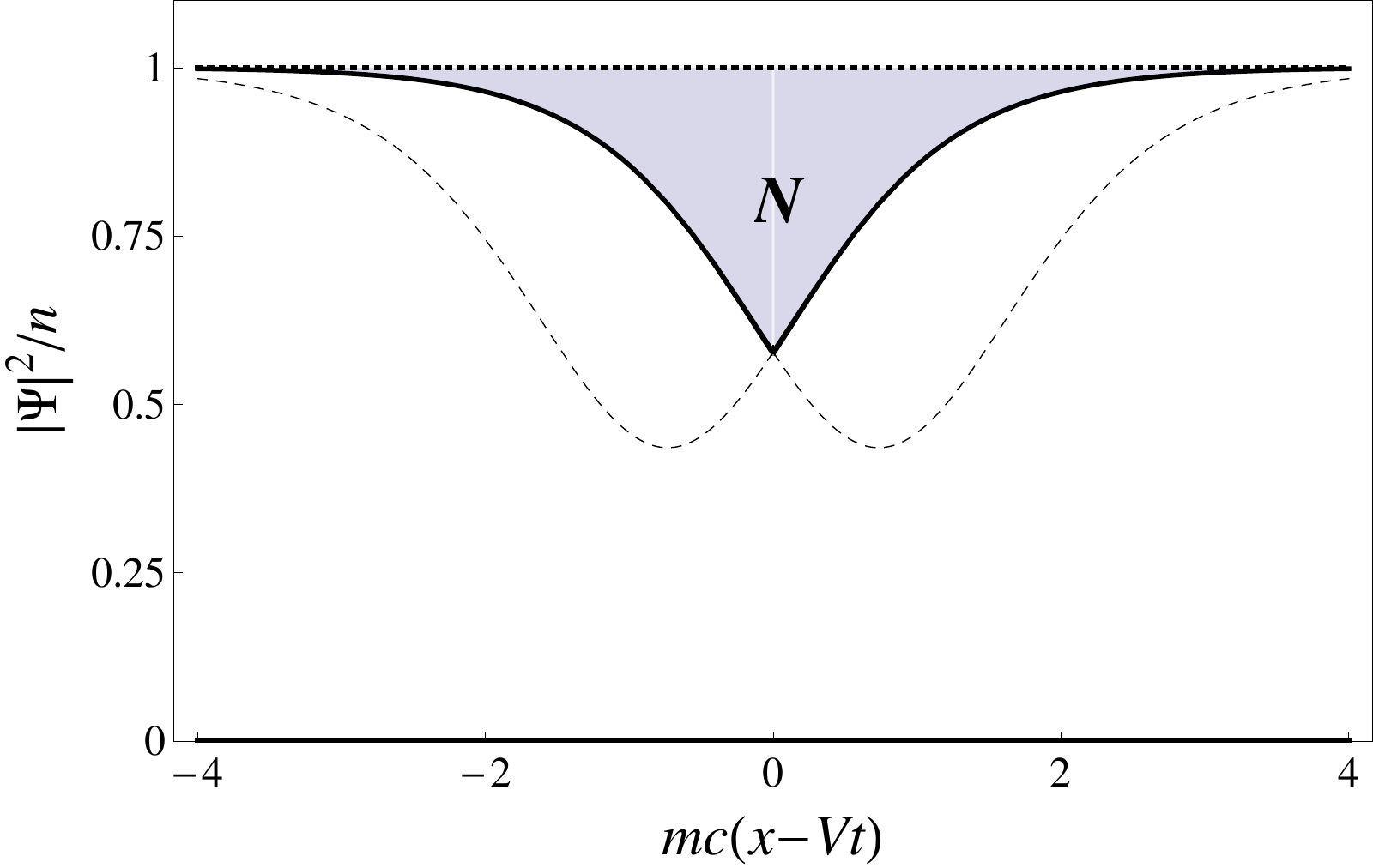}
\includegraphics[width=\columnwidth]{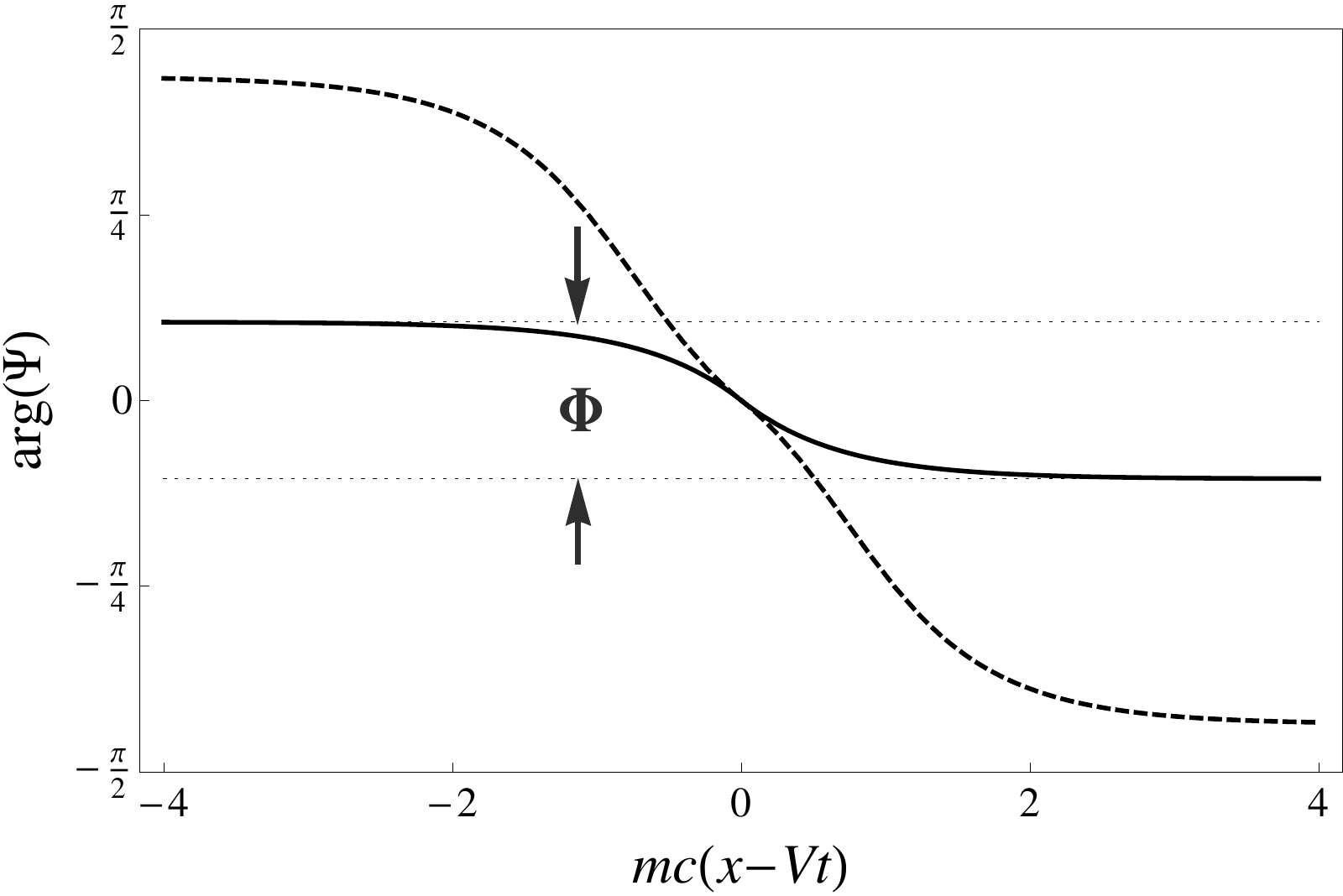}
\caption{Superfluid density (top panel) and phase (bottom panel) profiles in the presence of an impurity propagating with
velocity $V/c=0.66$ and coupling $G/c=0.25$ (solid lines). The repulsive impurity tends to deplete the host liquid, pushing away $N$ particles and inducing a phase drop $\Phi$ across it. The dashed lines
correspond to the double soliton solution Eq.(\ref{eq:wavefunction_no_impurity}) used to construct $\Psi(x-Vt)$.}
\label{fig:density}
\end{figure}

We may use these relations to compute the dressed impurity energy $E=\frac{1}{2}MV^2+E_{\mathrm{d}}$ and momentum $P=MV+P_{\mathrm{d}}$, where $E_{\mathrm{d}}$ and $P_{\mathrm{d}}$ are the corresponding depletion cloud contributions given by
\begin{eqnarray}
\label{eq:energy}
\nonumber E_{\mathrm{d}}&=&\int dx\left[ \frac{|\partial_x\Psi|^2}{2m}+\frac{g}{2}(n-|\Psi|^2)^2\right]+G|\Psi(X)|^2\\
&=&\frac{4}{3}nc\,
  \sin^{3}\frac{\Phi_{s}}{2}\left[1-\frac{3}{4}z-\frac{1}{4}z^{3}\right],\\
\label{eq:mom}	
\nonumber P_{\mathrm{d}}&=&\mathrm{Im}\int dx \left[\Psi^{*}\partial_x\Psi\right]-n\arg\Psi|^{x=\infty}_{x=-\infty}\\
&=&n\Phi-mNV.
\end{eqnarray}
Here we introduced the number of expelled particles $N=\frac{2n}{mc}\,\sin\frac{\Phi_s}{2}(1-z)$. Written in this way, the equilibrium expressions $E(\Phi+2\pi j)=E(\Phi)$ and $P(\Phi+2\pi j)=2\pi n j+P(\Phi)$ are functions of the total phase drop $\Phi$ across the impurity, whose functional forms depend only on the parameter $G/c$. Alternatively, one may solve for $\Phi=\Phi(P)$ with the help of Eq.~(\ref{eq:mom}). Substituting the resulting relation into the energy function, $E(\Phi(P))$, one obtains the periodic dispersion relation $E(P)=E(P+2\pi n)$.

\begin{figure}[t]
\centering
\includegraphics[width=\columnwidth]{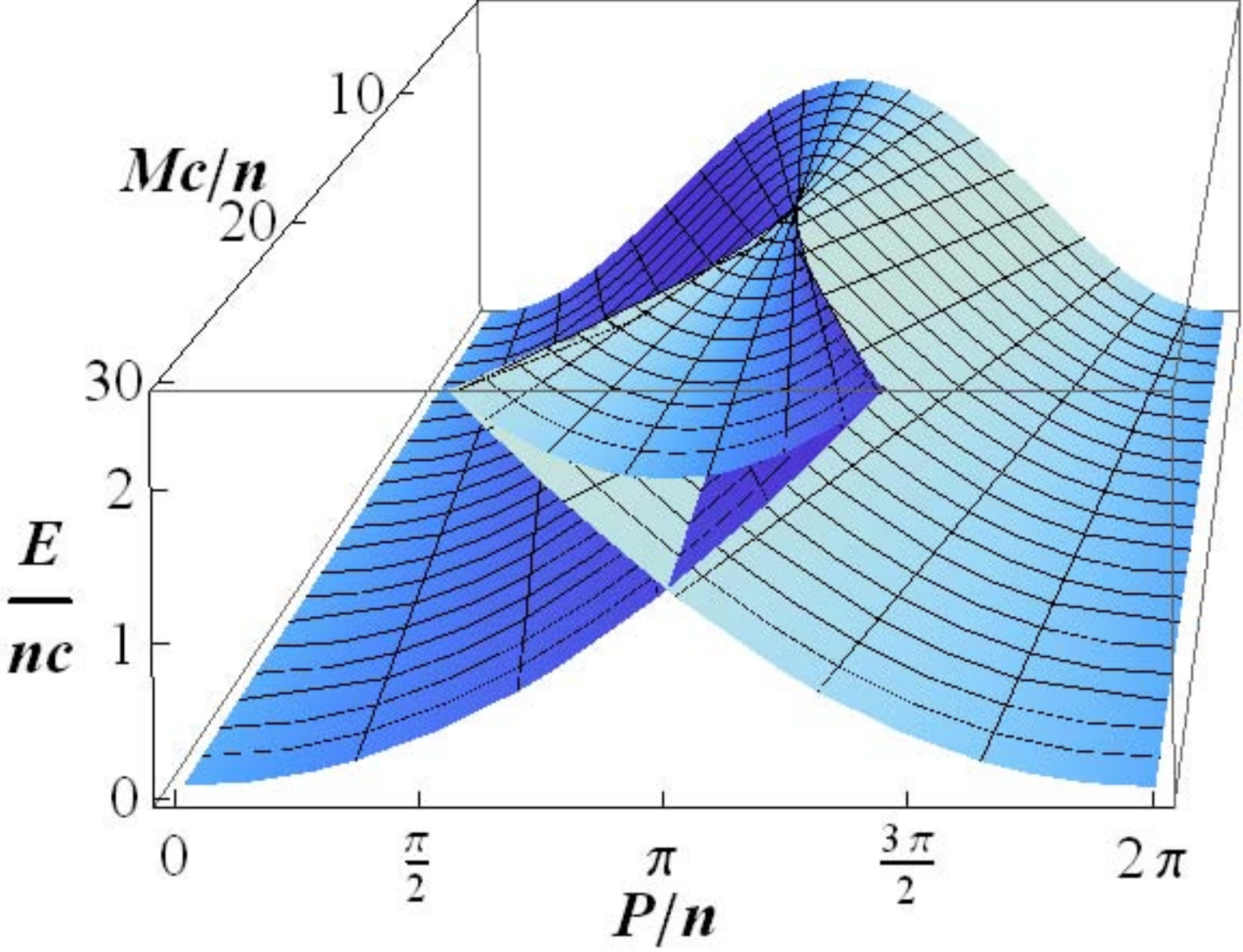}
\includegraphics[width=\columnwidth]{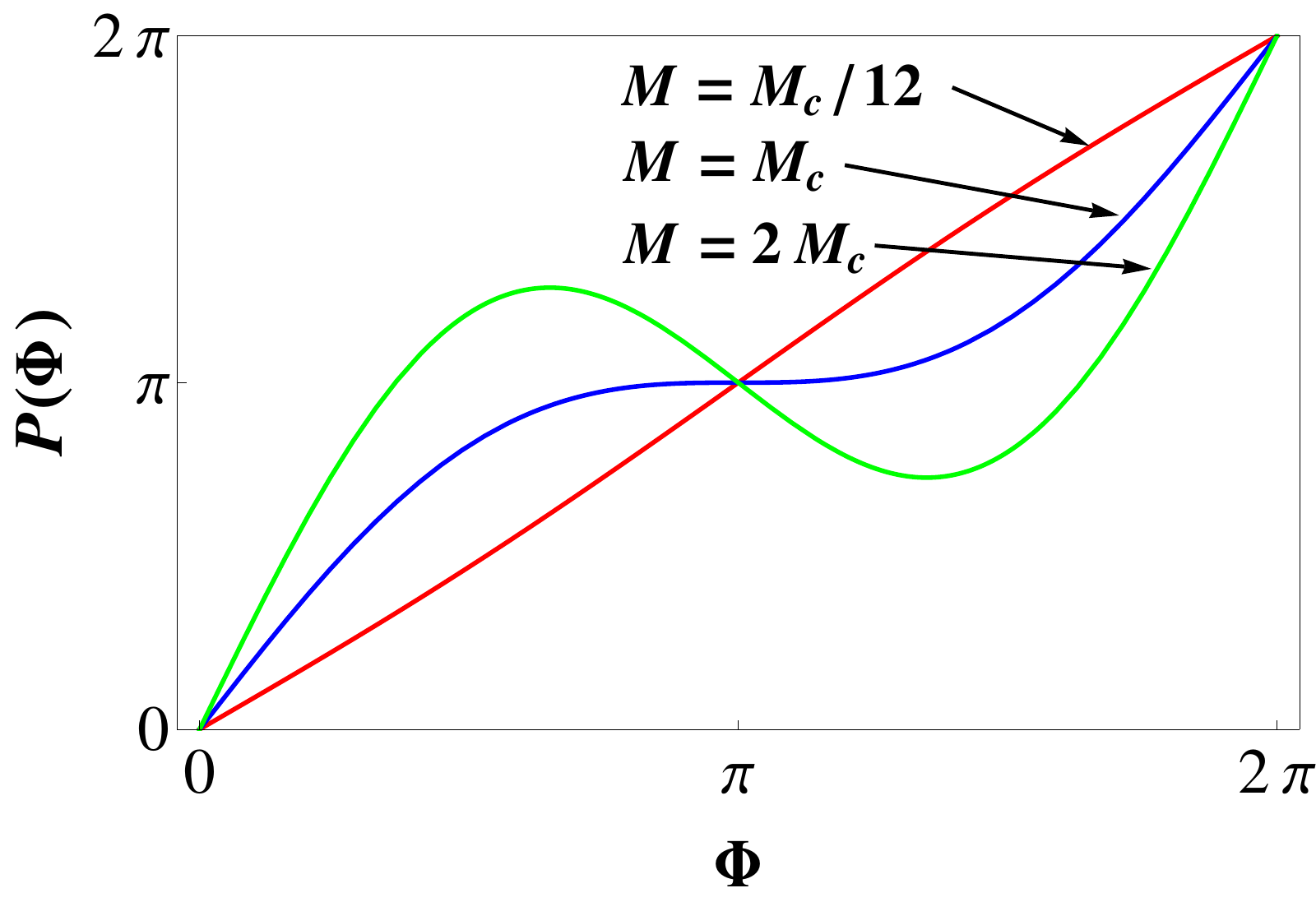}
\caption{Top Panel: Variation of the impurity dispersion across the $M=M_c$ transition, exhibiting the swallowtail catastrophe. Here $V_c/c=0.1$, $K=10\pi$ and $M_c =12n/c$ (see Eq.~(\ref{eq:M_c})). Bottom panel: Momentum $P(\Phi)$ for different masses, see Eq.~(\ref{eq:P}). For $M>M_c$ there persists two extrema which merge and disappear at $M=M_c$.}
\label{fig:energy}
\end{figure}
\begin{figure}[t]
\centering
\includegraphics[width=\columnwidth]{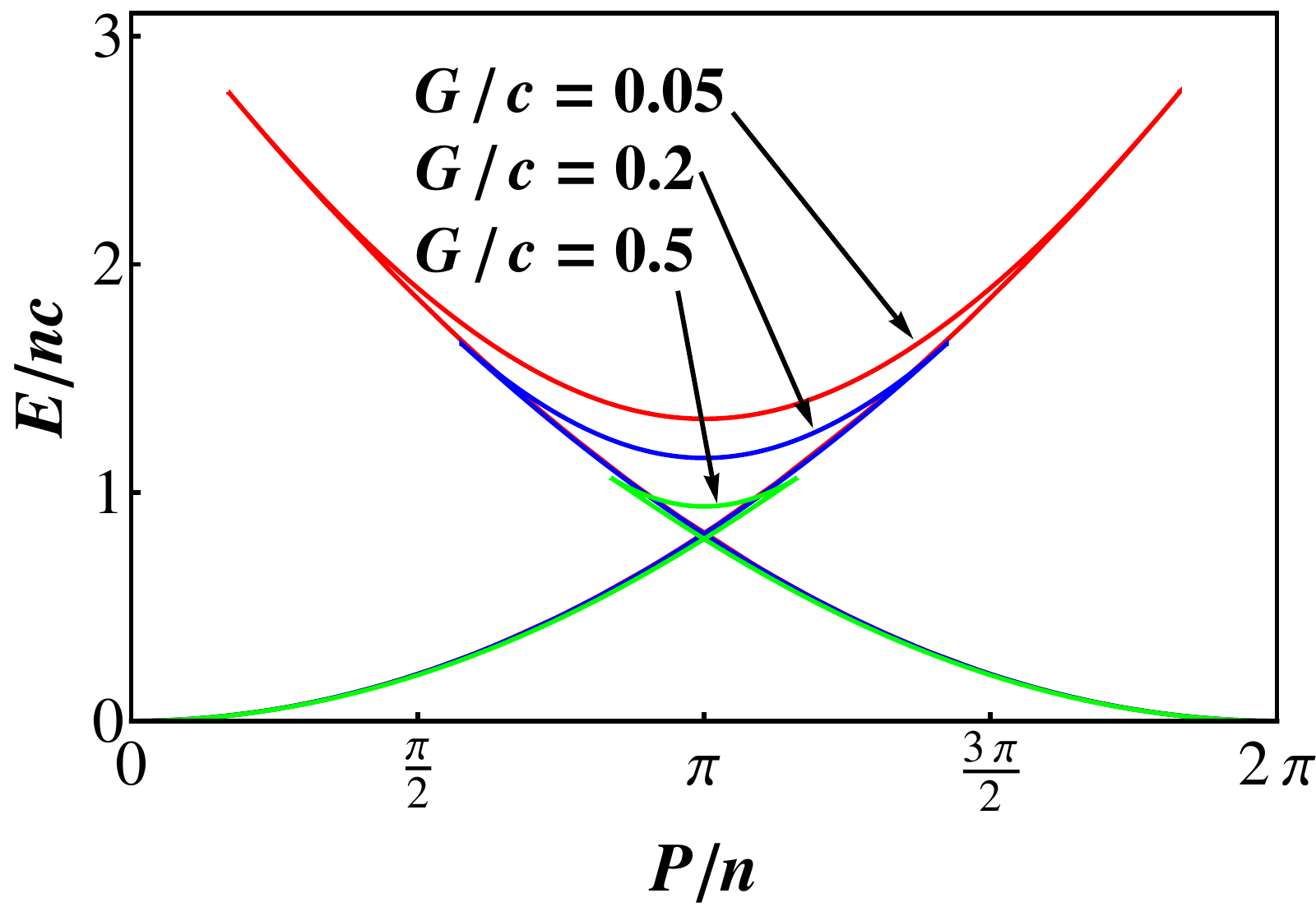}
\includegraphics[width=\columnwidth]{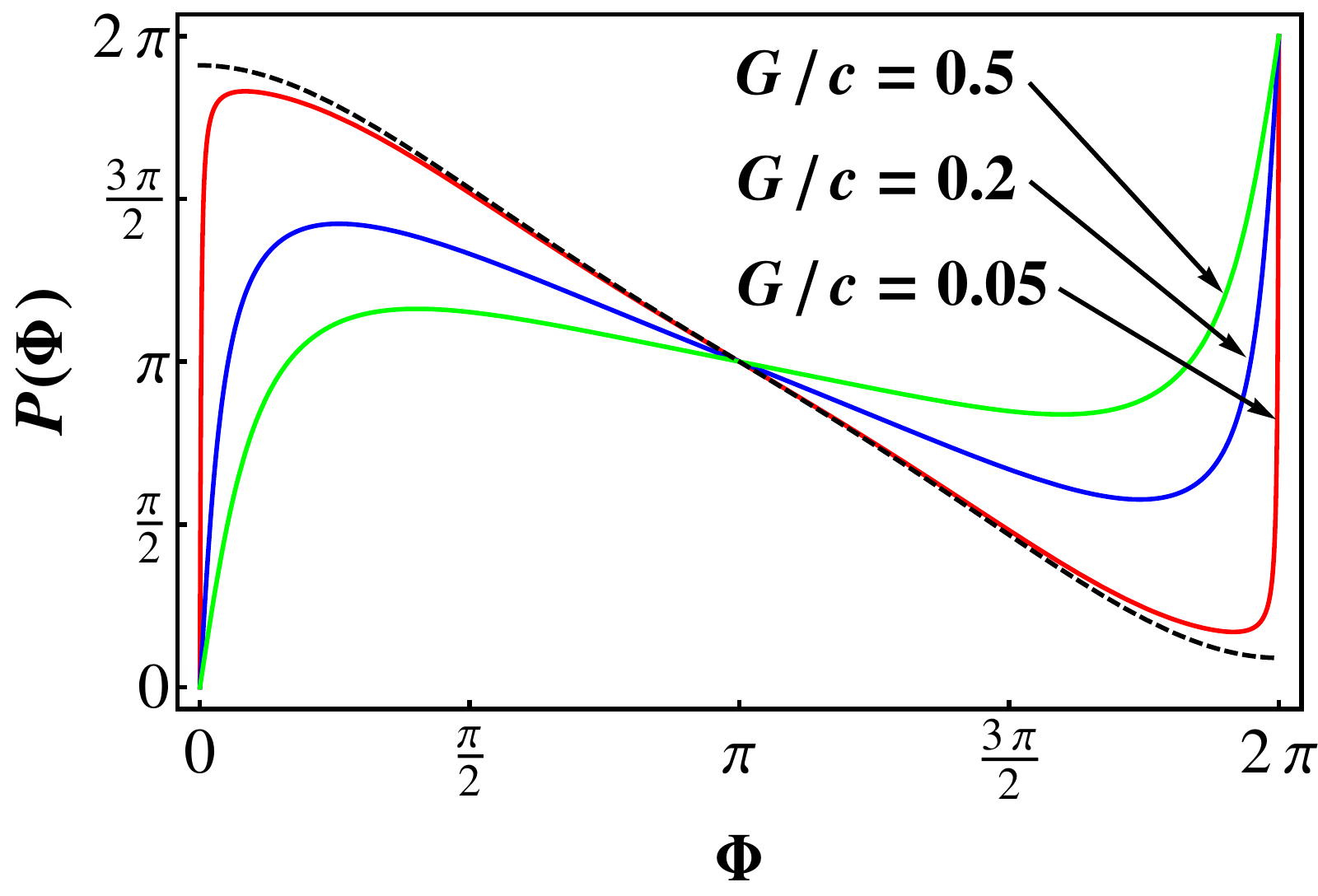}
\caption{Top panel: Impurity dispersion at weak coupling with parameters $M/m=60$, $K=10\pi$, $G/c=0.5$ (green), $G/c=0.2$ (blue) and $G/c=0.05$ (red). Bottom panel: Momentum $P$ as a function of $\Phi$ having the same parameters listed above. The $G\to0$ limit is well approximated by the gray soliton configuration (dashed): $P=n(\Phi-\mathrm{sin}\Phi)+Mc\,\mathrm{cos}\frac{\Phi}{2}$, except near $\Phi=2\pi j$ for integer $j$, where the momentum is necessarily given by $P=2\pi n j$.}
\label{fig:energy_weak_coupling}
\end{figure}

The above point description of the impurity is meaningful only if its effective mass is sufficiently large. This can be achieved either by a large bare impurity mass or strong coupling to the liquid. The latter tends to localize the impurity near the minimum of the self-induced depletion cloud density, see Fig.~\ref{fig:density}, while the former leads to a small de Broglie wavelength. To allow ourselves the generality of discussing a wide range of impurity masses, and for simplicity of illustration, we momentarily focus on the regime of strong coupling $G\gg c$.

\emph{Strong impurity coupling} -- We
determine the dependence of $z$ and $\Phi_\mathrm{s}$ on $\Phi$ to leading order in $c/G$. From Eqs.~(\ref{eq:z1}),~(\ref{eq:z2}) we find $z(\Phi)\simeq\frac{c}{G}\textrm{cos}^{2}\frac{\Phi}{2}$ and $\Phi_{s}(\Phi)\simeq\pi-\frac{c}{G}\sin\Phi$ and hence,
\begin{equation}
P=n\Phi+\left(M-\frac{2n}{c}\right) V_c\mathrm{sin}\Phi.
\label{eq:P}
\end{equation}
Here the critical velocity is given by $V_c=\frac{c^2}{2G}\ll c$ \cite{Hakim_PhysRevE.55.2835,Gunn_Taras_PhysRevB.60.13139}, and the $\Phi$ dependence of the velocity acquires the Josephson form: $V(\Phi)=V_c \mathrm{sin}\Phi$. The appearance of the critical mass occurs when $P=P(\Phi)$ is satisfied by multiple values of $\Phi(P)$. The functional form of $P(\Phi)$ for $M>M_c$ takes an "$\mathcal{N}$" shape, implying the existence of two extrema in the range $\Phi\in(0,2\pi)$, see Fig.~\ref{fig:energy}. As $M$ crosses $M_c$ from above, these extrema merge and disappear at $P=n\Phi=\pi n$. This observation allows to derive the critical mass within the above framework, 
\begin{equation}
\label{eq:M_c}
\frac{\partial P}{\partial \Phi}\Big|_{\Phi=\pi}=0\implies M_c=\frac{2n}{c}\left(1+\frac{G}{c}\right)\approx \frac{mK}{\pi}\frac{c}{V_c},
\end{equation}
in agreement with the parametric dependence Eq.~(3) of the main text.

\emph{Weakly coupled, heavy impurity} -- The case of weak impurity coupling does not lend itself to convenient analytical analysis. For given $G/c<1$, we thus solve the boundary Eqs.~(\ref{eq:z1}), (\ref{eq:z2}) numerically to produce the dispersion law $E(P)$, see Fig.~\ref{fig:energy_weak_coupling}. For comparison to the strong coupling case, we also plot $P(\Phi)$ for various couplings $G/c<1$. In the limit $G\to0$, the impurity is dressed by an essentially unperturbed gray soliton (\emph{i.e.}, the two solitons of Fig.~\ref{fig:density} are separated by a very small distance). This approximation breaks down near $\Phi=2\pi j$ for integer $j$ since we must have $P=2\pi n j$. Consequently, there is a sharp deviation from the soliton configuration (dashed line in Fig.~\ref{fig:energy_weak_coupling}) near these points, making the functional form of $P(\Phi)$ difficult to characterize analytically. As a result of the numerical analysis, we see that even in the regime of weak coupling the appearance of multiple momenta extrema give rise to the observed swallowtail catastrophe of the energy dispersion and is thus not restricted to the Josephson limit.

\emph{Quantum tunneling and non-analytic drift} --  We wish to calculate the tunneling-averaged energy drop $\langle\varepsilon\rangle$ as a function of the (small) force $F=P/t$ using the momentum dependent tunneling rate $\Gamma(P)$. To this end we establish the validity of the result (7) in the main text. 

The tunneling rate $\Gamma(\bar\tau)$ is given by the exponentiated action $S(\bar\tau)$ evaluated on the trajectory where the system tunnels for an imaginary time $\bar\tau$ from one minimum to the other. For $\alpha>1$ where tunneling is an irrelevant perturbation, we may safely neglect higher order processes involving tunneling back and forth between the minima and focus on the single tunneling event. Starting from Eqs.~(5),~(6) of the main text (after the Wick rotation $t\to-i\tau$) one integrates the imaginary time Luttinger liquid action with linear coupling to arrive at the tunneling action
\begin{equation}
\label{eq:action1}
S(\bar{\tau})=-\frac{1}{2}\int\mathrm{d}x\mathrm{d}x^{\prime}\mathrm{d}\tau\mathrm{d}\tau^{\prime}\vec{J}_{x,\tau}^{\dagger}(\bar{\tau})\hat{D}(x-x^{\prime},\tau-\tau^{\prime})\vec{J}_{x^{\prime},\tau^{\prime}}(\bar{\tau})
\end{equation}
where $\vec{J}^\dagger_{x,\tau}(\bar\tau)=\delta(x-X(\tau))(\dot\Phi,\dot N)=\delta(x-X(\tau))[\delta(\tau)-\delta(\tau-\bar\tau)](\Delta\Phi,\Delta N)$ describes the tunneling trajectory between minima having configurations which differ in phase by $\Delta\Phi$ and number of expelled particles $\Delta N$ and $\hat{D}(x,\tau)$ is the matrix imaginary time phononic propagator. The latter is determined by inverting the matrix operator of the quadratic Luttinger liquid action, Eq.~(5) of the main text, in the form $S_{\mathrm{LL}}=-\frac{1}{2}\int \mathrm{d}x \mathrm{d}\tau\vec\chi^\dagger\hat{D}^{-1}\vec\chi$, where $\vec\chi^\dagger=(\vartheta/\pi,\varphi)$. 

Substituting the above tunneling trajectories into Eq.~(\ref{eq:action1}) with the imaginary time propagator
\begin{equation}
\label{eq:propagator} 
\hat D(x,\tau)=\frac{1}{4\pi}\left[\begin{array}{cc}
\frac{K}{\pi}\mathrm{ln}\left(\frac{c^{2}\tau^{2}+x^{2}}{l^{2}}\right) & \mathrm{ln}\left(\frac{c\tau-ix}{c\tau+ix}\right)\\
\mathrm{ln}\left(\frac{c\tau-ix}{c\tau+ix}\right) & \frac{\pi}{K}\mathrm{ln}\left(\frac{c^{2}\tau^{2}+x^{2}}{l^{2}}\right)\end{array}\right],
\end{equation}
we find $S(\bar\tau)=(1+\alpha)\mathrm{ln}(c\bar\tau/l)$ where  $l\gtrsim(mc)^{-1}$ is a short distance cut-off beyond which the point-particle description of the depleton breaks down and $\alpha=2K\left[\left(\frac{\Delta\Phi}{2\pi}\right)^2+\left(\frac{\Delta N}{2K}\right)^2\right]-1$. The $X$ dependence of the action is neglected because it enters as $X(\tau)-X(\tau^\prime)\approx V(\tau-\tau^\prime)\ll c(\tau-\tau^\prime)$.

The energy-dependent tunneling rate is found with the help of the Fourier transformation, $\Gamma(\varepsilon)\sim\int\mathrm{d}t e^{-S(it)}e^{i\varepsilon t}$, where the analytical continuation to real energy $i\varepsilon\to\varepsilon$ is accompanied by the Wick rotation $\tau\to it$. Substituting $S(\tau)$ given above and using the identity $\int \mathrm{d}t e^{i\varepsilon t}(it)^{-(1+\alpha)}=\frac{2\pi\Theta(\varepsilon)}{\Gamma(1+\alpha)}\varepsilon^\alpha$, we arrive at Eq.~(7) of the main text.

To compute $\langle\varepsilon\rangle$ one must know the probability to tunnel per unit time: $\dot{\mathcal{P}}(t)=\Gamma(t)\textrm{Exp}\left[-\int_{0}^{t}dt^{\prime}\Gamma(t^{\prime})\right]$, which follows from the assumption that the process is Markovian. By parameterizing the average over time by an average over momenta one finds
\begin{eqnarray}
\label{eq:ep_avg}
\langle\varepsilon\rangle &=&\int_{n\pi}^{P_{t}}dP\varepsilon(P)\frac{\Gamma(P)}{F}\textrm{Exp}\left[-\int_{n\pi}^{P}dP^{\prime}\frac{\Gamma(P^{\prime})}{F}\right]\\
&=&\int_{0}^{F_{\pi}/F}dx\varepsilon(P(x))e^{-x}.
\end{eqnarray}
Here we introduced the dimensionless variable $x(P)=\int_{n\pi}^{P}dP^{\prime}\Gamma(P^{\prime})/F$ along with $F_{\pi}=\int_{n\pi}^{P_{t}}dP^{\prime}\Gamma(P^{\prime})$.  Due to the exponential decay of the kernel in Eq.~(\ref{eq:ep_avg}), the integral is cut at $x_c=\textrm{min}\{1,\, F_\pi/F\}$. As a result, one obtains the approximation $\langle\varepsilon\rangle\sim x_c\varepsilon(P_c)$, where $P_c$ satisfies $x_c=\int_{n\pi}^{P_c}dP\Gamma(P)/F$. For $F\ll F_\pi$, we have $x_c=1$ and hence $P_c$ may be solved by writing $P_c=n\pi+\delta P$ for small, yet undetermined, $\delta P$. Expanding $\Gamma$ near $n\pi$ as $\Gamma=\Gamma_{\pi}[(P-n\pi)\mathcal{V}_c/mc^2]^{\alpha}$, we find $\delta P=\frac{mc^2}{\mathcal{V}_c} (F/F_c)^{\frac{1}{1+\alpha}}$ and $F_c=\frac{\Gamma_\pi mc^2}{\mathcal{V}_c(1+\alpha)}\approx\frac{\Gamma_\pi Mc}{2K^2}$ for large $M$. Thus,
\begin{equation}
\label{eq:drift}
\langle\varepsilon\rangle=mc^2\left(\frac{F}{F_c}\right)^{\frac{1}{1+\alpha}}\implies V_{\mathrm{D}}=\frac{c}{K}\left(\frac{F}{F_c}\right)^{\frac{1}{1+\alpha}},
\end{equation}
establishing Eq.~(9) of the main text.

\end{document}